# Amplitude Quantization for Type-2 Codebook Based CSI Feedback in New Radio System

Honglei Miao and Markus D. Mueck and Michael Faerber
Intel Deutschland GmbH
Am Campeon 10-12, 85579, Neubiberg, Germany
{honglei.miao, markus.mueck, michael.faerber}@intel.com

*Abstract*—In 3GPP new radio system, two types of codebook, namely Type-1 and Type-2 codebook, have been standardized for the channel state information (CSI) feedback in the support of advanced MIMO operation. Both types of codebook are constructed from 2-D DFT based grid of beams, and enable the CSI feedback of beam selection as well as PSK based co-phase combining between two polarizations. Moreover, Type-2 codebook based CSI feedback reports the wideband and subband amplitude information of the selected beams. As a result, it is envisioned that more accurate CSI shall be obtained from the Type-2 codebook based CSI feedback so that better precoded MIMO transmission can be employed by the network. To reduce the CSI feedback signaling, 1 bit based subband amplitude with only two quantization levels is supported in combination to 3 bits based wideband amplitude feedback. Typically, wideband amplitude shall be calculated as the linear average amplitude of the beam over all subbands. However, due to the coarse subband amplitude quantization, it has been observed in case of joint wideband and subband amplitude feedback, the average based wideband amplitude can lead to a large amplitude quantization errors. In this paper, we study two methods for joint wideband and subband amplitude calculations. Specifically, both optimal and sub-optimal methods are proposed. The optimal method can achieve the minimum amplitude quantization errors at the cost of a relatively large computation complexity. And by virtue of a derived scaling factor, the sub-optimal method exhibits clearly smaller quantization error than the conventional linear average based method especially for the channel with large frequency selectivity.

## I. INTRODUCTION

In 3GPP new radio system, Type-1 and Type-2 codebook based channel state information (CSI) feedback have been standardized [1] [2] to support advanced MIMO transmission. Both types of codebooks are constructed from 2-D DFT based grid of beams, and enable the CSI feedback of beam selection as well as PSK based co-phase combining between two polarizations. In addition, Type-2 codebook based CSI feedback [2] [3] also reports the wideband (WB) and subband (SB) amplitude information of the selected beams. As a result, it is envisioned that more accurate CSI can be obtained from Type-2 codebook based CSI feedback so that better precoded MIMO transmission can be employed by the network. To achieve low CSI feedback signaling overhead, 1 bit based subband amplitude with only two quantization levels is specified in combination with 3 bits based wideband amplitude feedback. Specifically, SB amplitude feedback of 0 indicates the quantized SB channel power to be the half of the signaled wideband power while SB amplitude feedback of 1 informs the quantized SB channel power to be equal to the signaled wideband channel power.

Typically, wideband amplitude shall be calculated as the linear average amplitude of the beam over all subbands. However, due to the coarse subband amplitude quantization, it has been observed in case of joint wideband and subband (Joint-WB-and-SB) amplitude feedback, the linear average based wideband amplitude can lead to a large amplitude quantization errors. In this paper, we develop two methods, namely optimal method and sub-optimal method, for joint wideband and subband amplitude calculations. Specifically, the optimal method can achieve the minimum overall amplitude quantization errors at the cost of a relatively large computation complexity. With a derived scaling factor, the sub-optimal method exhibits smaller quantization error than the conventional linear average based method especially for the channel amplitude with large frequency selectivity.

The paper is organized as follows. In Section II Type-2 codebook based CSI feedback and quantization error analysis of linear average based WB amplitude for Joint-WB-and-SB PMI feedback are presented. In Section III optimal method to achieve minimum amplitude quantization error is detailed. In Section IV sub-optimal method with reduced computation complexity is described. In Section V simulation results of developed methods are given. Finally Section VI concludes the paper.

## II. PROBLEM STATEMENT

In this section, we present Type-2 codebook based CSI feedback and quantization error analysis of linear average based WB amplitude for Joint-WB-and-SB PMI feedback.

### A. CSI Feedback with Type-2 Codebook

According to [2] [3], Type-2 codebook based CSI feedback enables more explicit channel feedback than Type-1 based CSI feedback in the sense that both beam direction and amplitude are reported by the UE in the CSI feedback. As a result, channel can be characterized with more accurate spatial and amplitude

With Type-2 codebook, up to two layers of clustered beams can be signalled by the UE. For rank 1, precoder matrix indicator (PMI) feedback $W = \begin{bmatrix} \widetilde{w}_{0,0} \\ \widetilde{w}_{1,0} \end{bmatrix}$, $W$ is normalized to 1. For rank 2, PMI feedback $W = \begin{bmatrix} \widetilde{w}_{0,0} & \widetilde{w}_{0,1} \\ \widetilde{w}_{1,0} & \widetilde{w}_{1,1} \end{bmatrix}$, columns of $W$ are normalized to $\frac{1}{\sqrt{2}}$.. And $\widetilde{w}_{r,l}$ represents weighted combination of $L$ beams, and can be expressed as follows.

$$\widetilde{w}_{r,l} = \sum_{i=0}^{L-1} b_{\theta_1^{(i)} \theta_2^{(i)}} \cdot a_{r,l,i}^{(WB)} \cdot a_{r,l,i}^{(SB)} \cdot c_{r,l,i} \quad (1)$$

where the value of $L$ is configurable, $L \in \{2,3,4\}$, defines the number of beams per layer, $b_{\theta_1^{(i)} \theta_2^{(i)}}$ is an oversampled 2D DFT beam, $\theta_1^{(i)}$ and $\theta_2^{(i)}$ refer to the beam index in horizontoal and vertical domain, respectively, $r = 0,1$ denotes the polarization direction, $l = 0,1$ the layer index. $a_{r,l,i}^{(WB)}$ stands for wideband beam amplitude scaling factor for beam i and on polarization $r$ and layer $l$, and $a_{r,l,i}^{(SB)}$ refers to subband beam amplitude scaling factor for beam i and on polarization $r$ and layer $l$. $c_{r,l,i}$ denotes beam phase combining coefficient for beam $i$ and on polarization $r$ and layer $l$, and it can be configurable between QPSK (2 bits) and 8PSK (3 bits).

As described in [2] [3], amplitude coefficient vectors $a_l^{(1)}$ and $a_l^{(2)}$ are defined as follows:

$$a_l^{(1)} = (a_{0,l,0}^{(WB)}, \ldots, a_{0,l,L-1}^{(WB)}, a_{1,l,0}^{(WB)}, \ldots, a_{1,l,L-1}^{(WB)}), \quad (2)$$
$$a_l^{(2)} = (a_{0,l,0}^{(SB)}, \ldots, a_{0,l,L-1}^{(SB)}, a_{1,l,0}^{(SB)}, \ldots, a_{1,l,L-1}^{(SB)}). \quad (3)$$

And $a_{l,i}^{(1)}$ and $a_{l,i}^{(2)}$ denote the $i$th entry of vectors $a_l^{(1)}$ and $a_l^{(2)}$, respectively. The WB amplitude indicator $k_l^{(1)}$ and the SB amplitude indicator $k_l^{(2)}$ are defined as

$$k_l^{(1)} = (k_{l,0}^{(1)}, k_{l,1}^{(1)}, \ldots, k_{l,2L-1}^{(1)}), \quad (4)$$
$$k_l^{(2)} = (k_{l,0}^{(2)}, k_{l,1}^{(2)}, \ldots, k_{l,2L-1}^{(2)}), \quad (5)$$

where
$$k_{l,i}^{(1)} \in \{0,1,\ldots,7\}; k_{l,i}^{(2)} \in \{0,1\}$$

for $l = 1,\ldots,\upsilon$, where $\upsilon$ defines the number of layers, and $k_{l,i}^{(1)}$ and $k_{l,i}^{(2)}$ are indies of quantized value of $a_{l,i}^{(1)}$ and $a_{l,i}^{(2)}$, respectively. Specifically, $k_{l,i}^{(1)}$ of 3 bits ranges from 0 to 7, where 0 corresponds to zero amplitude and other values define 7 amplitude levels in the unit of decibel with 3dB increase step, and $k_{l,i}^{(2)}$ is defined by 1 bit, thereby 0 refers to -3dB and 1 for 0dB.

According to [2] [3], the PMI feedback supports configurable amplitude feedback mode between WB-Only mode where only $k_{l,i}^{(1)}$ are reported, and Joint-WB-and-SB mode where both $k_{l,i}^{(1)}$ and $k_{l,i}^{(2)}$ are reported. The method for PMI feedback calculation for WB-only and Joint-WB-and-SB modes is illustrated in Fig. 1.

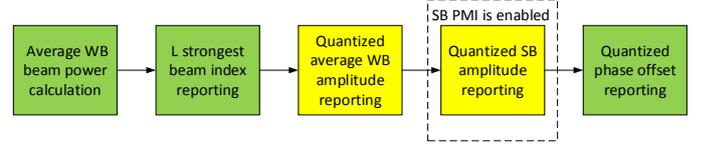

Fig. 1. Type-2 codebook based WB-Only and Joint-WB-and-SB PMI feedback

From Eq. (2), the WB average amplitude vector $p_l^{(1)}$ in decibel of $L$ selected beams in two polarizations can be expressed as follows

$$p_l^{(1)} = \left(p_{l,0}^{(WB)}, p_{l,1}^{(WB)}, \ldots, p_{l,2L-1}^{(WB)}\right) \quad (6)$$

where $p_{l,i}^{(WB)} = a_{\left\lfloor \frac{i}{L} \right\rfloor, l, \mod(i,L)}^{(WB)}$ defines the WB beam amplitude for the logical beam $i$ and layer $l$. It is noted that different polarizations of same physical beam are numbered as different logical beams.

Eight quantized amplitude levels $\bar{Q}_{l,i}^{(1)}, i = 0,1,2\ldots,7$, are defined as follows

$$\bar{Q}_{l,7}^{(1)} = \max_i p_{l,i}^{(WB)}, \quad (7)$$
$$\bar{Q}_{l,i}^{(1)} = \bar{Q}_{l,7}^{(1)} - (7-i) \cdot 10 \log 2, i = 0,1,\ldots,6 \quad (8)$$

Based on the above quantized amplitude level $\bar{Q}_{l,i}^{(1)}$, the quantized WB amplitude index $k_{l,i}^{(1)}$ can be calculated as follows
$$k_{l,i}^{(1)} = \left\{m \middle| \left|p_{l,i}^{(WB)} - \bar{Q}_{l,m}^{(1)}\right| = \min_n \left|p_{l,i}^{(WB)} - \bar{Q}_{l,n}^{(1)}\right|\right\}$$

In case of Joint-WB-and-SB PMI feedback, two quantized amplitude levels in decibel for each beam are defined as

$$\bar{Q}_{l,i,1}^{(2)} = \bar{Q}_{l,k_{l,i}^{(1)}}^{(1)}, \quad (9)$$
$$\bar{Q}_{l,i,0}^{(2)} = \bar{Q}_{l,i,1}^{(2)} - 10 \log 2 \quad (10)$$

Let $S$ define the number of supported SBs, and $p_{l,i,s}^{(SB)}$ the average amplitude in decibel of beam $i$ at SB $s, s = 0,1,\ldots,S-1$. For each SB $s$, the quantized SB amplitude index $k_{l,i,s}^{(2)}$ can be obtained as follows

$$k_{l,i,s}^{(2)} = \left\{m \middle| \left|p_{l,i,s}^{(SB)} - \bar{Q}_{l,i,m}^{(2)}\right| = \min_{n \in \{0,1\}} \left|p_{l,i,s}^{(SB)} - \bar{Q}_{l,i,n}^{(2)}\right|\right\}. \quad (11)$$

### B. Quantization Error of Joint-WB-and-SB PMI Feedback

It is observed from Section A that two-level SB amplitude quantizer for each beam is based on the WB amplitude of the beam over all SBs. For example, as shown in Fig. 2, given two SBs for each beam, and unquantized amplitude values in two

SBs of a beam are (0.5, 1), the linear average amplitude value of the beam over two SBs is 0.75, which can be signaled as the WB amplitude value. And the two quantization levels of the SB amplitude quantizer based on WB amplitude of linear average value are [0.375, 0.75]. As a result, the quantized amplitudes of two SBs are {0.375 0.75}.

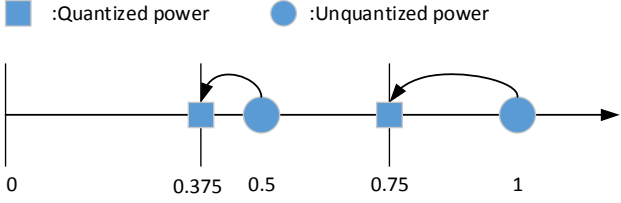

Fig. 2. Average WB amplitude based 2-level SB amplitude quantizer

The Root Normalized Squred Quantized Error (RNSQE) in Fig. 2 can be calculated as

$$\varepsilon_Q = \sqrt{\frac{|0.5-0.375|^2+|1-0.75|^2}{0.5^2+1}} = 0.25. \quad (12)$$

It is observed from above that the linear average WB amplitude based 2-level SB amplitude quantizer can cause quite large RNSQE, e.g., 25% in the above example. This leads to a question how to improve the 2-level SB amplitude quantizer to achieve smaller normalized amplitude quantization error for Joint-WB-and-SB PMI feedback.

### III. OPTIMAL METHOD

It is shown in Section II that WB amplitude based on linear average of SB amplitudes can cause large quantization errors for Joint-WB-and-SB PMI feedback. In this section, we present an optimal method for calculating the WB beam amplitude, based on which the minimum normalized amplitude quantization error can be achieved for Joint-WB-and-SB PMI feedback.

*A. Algorithm Derivation*

Let $\boldsymbol{p}_{l,i}^{(SB)} = \left(p_{l,i,0}^{(SB)}, p_{l,i,1}^{(SB)}, \ldots, p_{l,i,S-1}^{(SB)}\right)$ define the observed SB amplitude vector for the beam $i$ and layer $l$. $\boldsymbol{r}_{l,i} = \left(r_{l,i,0}, r_{l,i,1}, \ldots, r_{l,i,S-1}\right)$ defines the quantized SB amplitude vector for the beam $i$ and layer $l$, where $r_{l,i,s}$ is the quantized value of $p_{l,i,s}^{(SB)}$ based on 2-level quantizer determined by $p_{l,i}^{(WB)}$ as follows.

$$r_{l,i,s} = \begin{cases} 1, & p_{l,i,s}^{(SB)} \geq \frac{3}{4}p_{l,i}^{(WB)} \\ \frac{1}{2}, & p_{l,i,s}^{(SB)} < \frac{3}{4}p_{l,i}^{(WB)} \end{cases} \quad (13)$$

Given observed SB amplitude vector $\boldsymbol{p}_{l,i}^{(SB)}$, unlike linear average based WB amplitude $p_{l,i}^{(Lin-WB)} = \text{mean}\left(\boldsymbol{p}_{l,i}^{(SB)}\right)$, the optimal WB amplitude $p_{l,i}^{(Opt-WB)}$ shall be calculated to minimize total squared quantization error as follows.

$$p_{l,i}^{(Opt-WB)} = \min_{p_{l,i}^{(WB)}} \left(\sum_{s=0}^{S-1}(p_{l,i}^{(WB)}r_{l,i,s} - p_{l,i,s}^{(SB)})^2\right), \quad (14)$$

where $r_{l,i,s}$ is calculated according to Eq. (13).

To solve the optimal problem in Eq. (14), without loss of generality, one can assume $p_{l,i,0}^{(SB)} \leq p_{l,i,1}^{(SB)} \leq \ldots \leq p_{l,i,S-1}^{(SB)}$, the feasible region of $p_{l,i}^{(WB)}$ can be divided into following $S+1$ nonoverlapped sub-regions:

$$R_0 = \left\{p_{l,i}^{(WB)} \leq \tfrac{4}{3}p_{l,i,0}^{(SB)}\right\},$$
$$R_n = \left\{\tfrac{4}{3}p_{l,i,n-1}^{(SB)} \leq p_{l,i}^{(WB)} \leq \tfrac{4}{3}p_{l,i,n}^{(SB)}\right\}, n = 1 \ldots S-1$$
$$R_S = \left\{\tfrac{4}{3}p_{l,i,S-1}^{(SB)} \leq p_{l,i}^{(WB)}\right\},$$

Accordingly quantized SB amplitude vector $\boldsymbol{r}_{l,i}$ can be expressed as

$$\boldsymbol{r}_{l,i} = \begin{cases} \boldsymbol{r}^{(0)} = (1,1,\ldots,1), & p_{l,i}^{(WB)} \in R_0 \\ \boldsymbol{r}^{(1)} = \left(\tfrac{1}{2},1,\ldots,1\right), & p_{l,i}^{(WB)} \in R_1 \\ \vdots & \vdots \\ \boldsymbol{r}^{(S)} = \left(\tfrac{1}{2},\tfrac{1}{2},\ldots,\tfrac{1}{2}\right), & p_{l,i}^{(WB)} \in R_S \end{cases} \quad (15)$$

With the above definitions, the optimization problem in Eq. (14) can be reformulated as the following problem

$$p_{l,i}^{(Opt-WB)} = \min_{n=0,\ldots S} \left(\min_{p_{l,i}^{(WB)} \in R_n} g^{(n)}\left(p_{l,i}^{(WB)}, \boldsymbol{r}^{(n)}\right)\right), \quad (16)$$

where

$$g^{(n)}\left(p_{l,i}^{(WB)}, \boldsymbol{r}^{(n)}\right) = \sum_{s=0}^{S-1}\left(p_{l,i}^{(WB)} r_s^{(n)} - p_{l,i,s}^{(SB)}\right)^2. \quad (17)$$

Let $p_{l,i}^{(n,*)}$ define the unconstrained minimizor of $g^{(n)}\left(p_{l,i}^{(WB)}, \boldsymbol{r}^{(n)}\right)$, and it can be obtained as follows

$$p_{l,i}^{(WB,n,*)} = \frac{\sum_{s=0}^{S-1} p_{l,i,s}^{(SB)} r_s^{(n)}}{\sum_{s=0}^{S-1}\left(r_s^{(n)}\right)^2}. \quad (18)$$

As a result, the solution to the constrained minization $\min_{p_{l,i}^{(WB)} \in R_n} g^{(n)}\left(p_{l,i}^{(WB)}, \boldsymbol{r}^{(n)}\right)$ can be expressed as

$$\bar{p}_{l,i}^{(WB,n)} = \begin{cases} p_{l,i}^{(WB,n,*)}, & p_{l,i}^{(WB,n,*)} \in R_n \\ \tfrac{4}{3}p_{l,i,0}^{(SB)}, & p_{l,i}^{(WB,0,*)} \notin R_0, n=0 \\ \tfrac{4}{3}p_{l,i,S-1}^{(SB)}, & p_{l,i}^{(WB,S,*)} \notin R_S, n=S \\ \min_{p_{l,i}^{(WB)} \in \left\{\tfrac{4}{3}p_{l,i,n-1}^{(SB)}, \tfrac{4}{3}p_{l,i,n}^{(SB)}\right\}} g^{(n)}\left(p_{l,i}^{(WB)}, \boldsymbol{r}^{(n)}\right), & p_{l,i}^{(WB,n,*)} \notin R_n, n=1,\ldots,S-1 \end{cases}$$

(19)

The optimal WB amplitude $p_{l,i}^{(\text{Opt-WB})}$, i.e., the solution to Eq. (14), can be obtained as

$$p_{l,i}^{(\text{Opt-WB})} = \left\{ \bar{p}_{l,i}^{(\text{WB},k)} \Big| g^{(k)}(\bar{p}_{l,i}^{(\text{WB},k)}, r^{(k)}) = \min_{n=0,\ldots S}\left(g^{(n)}(\bar{p}_{l,i}^{(\text{WB},n)}, r^{(n)})\right) \right\}, \quad (20)$$

It should be noted that above method focuses on the optimal WB amplitude calculation for one beam. In case of multiple beam feedback given in Eq. (6), the same prinple used for one beam calcuation can be straightforwardly extended. For the sake of compactness, this extension is omitted in this paper.

*B. Numerical Example*

In this section, a simple numerical example is given to demonstrate the quantization error performance of the presented optimal method compared to the conventional linear average based WB amplitude.

For the example in Section II.B, where two observed SB amplitudes of a beam are $(0.5, 1)$, three sub-regions are calculaed as

$$R_0 = \left\{ p_{l,i}^{(\text{WB})} \leq \tfrac{2}{3} \right\},$$
$$R_1 = \left\{ \tfrac{2}{3} \leq p_{l,i}^{(\text{WB})} \leq \tfrac{4}{3} \right\},$$
$$R_2 = \left\{ \tfrac{4}{3} \leq p_{l,i}^{(\text{WB})} \right\}.$$

According to Eq. (18), $p_{l,i}^{(\text{WB},0,*)} = \tfrac{3}{4}$, $p_{l,i}^{(\text{WB},1,*)} = 1$, and $p_{l,i}^{(\text{WB},2,*)} = \tfrac{3}{2}$. And based on Eq. (19), $\bar{p}_{l,i}^{(\text{WB},0)} = \tfrac{2}{3}$, $\bar{p}_{l,i}^{(\text{WB},1)} = 1$, and $\bar{p}_{l,i}^{(\text{WB},2)} = \tfrac{3}{2}$. Moreover, according to Eq. (17), $g^{(0)}(\bar{p}_{l,i}^{(\text{WB},0)}, r^{(0)}) = \tfrac{17}{36}$, $g^{(1)}(\bar{p}_{l,i}^{(\text{WB},1)}, r^{(1)}) = 0$, and $g^{(2)}(\bar{p}_{l,i}^{(\text{WB},2)}, r^{(2)}) = \tfrac{1}{8}$.

Apparently, it is clear from Eq. (20) that $p_{l,i}^{(\text{Opt-WB})} = \bar{p}_{l,i}^{(\text{WB},1)} = 1$, this would lead to zero total squared quantization error which is obviously the minimum achievable error. In this case, the optimal WB amplitude $p_{l,i}^{(\text{Opt-WB})}$ is 1, which is the maximum SB amplitude instead of the linear average WB amplitude over two SBs.

## IV. SUB-OPTIMAL METHOD WITH REDUCED COMPLEXITY

It is clear from Section III that optimal WB amplitude calculation by Eq. (20) is a quite complicated procedure. It can be interesting to find a sub-optimal method with reduced complexity. In this section, a sub-optimal method with reduced complexity and reasonably small quantization error is provided.

*A. Algorithm Derivation*

Specifically, the sub-optimal method is designed to minimize the mean squared quantization error given a particular assumption that $r_{l,i,s}$ is random value with equal probability to be either ½ or 1, and independent from $p_{l,i}^{(\text{WB})}$. It is obvious that such assumption does not agree with the real situation. However, it enables the simplification of the calculation procedure, and can achieve reasonably good quantization error performance according to the simulation results in Section V.

Given the above assumption that $r_{l,i,s}$ is randam variable with equal probability to be either ½ or 1, and independent from $p_{l,i}^{(\text{WB})}$, the sub-optimal WB beam amplitude $p_{l,i}^{(\text{Sub-WB})}$ is calculated as follows

$$p_{l,i}^{(\text{Sub-WB})} = \min_{p_{l,i}^{(\text{WB})}} E\left( \sum_{s=0}^{S-1} (p_{l,i}^{(\text{WB})} r_{l,i,s} - p_{l,i,s}^{(\text{SB})})^2 \right) \quad (21)$$

The Eq. (21) can be further expressed as

$$p_{l,i}^{(\text{Sub-WB})} = \min_{p_{l,i}^{(\text{WB})}} \sum_{s=0}^{S-1} \left( \left( \tfrac{p_{l,i}^{(\text{WB})}}{2} - p_{l,i,s}^{(\text{SB})} \right)^2 + \left( p_{l,i}^{(\text{WB})} - p_{l,i,s}^{(\text{SB})} \right)^2 \right) \quad (22)$$

With some mathematical manipulation, the solution to Eq. (22) can be obtained as

$$p_{l,i}^{(\text{Sub-WB})} = \frac{6 \sum_{s=0}^{S-1} p_{l,i,s}^{(\text{SB})}}{5S} \quad (23)$$

It is shown from above solution that the sub-optimal WB amplitude shall be the linear average based WB amplitude scaled by 6/5 for Joint-WB-and-SB PMI feedback. The key points of the sub-optimal method stem from the assumption that quantized SB amplitude $r_{l,i,s}$ is random variable independent from $p_{l,i}^{(\text{WB})}$ and $p_{l,i,s-1}^{(\text{SB})}$. Such assumption relaxes the problem of finding optimal value for $p_{l,i}^{(\text{WB})}$ at the cost of non-optimal RNSQE.

*B. Numerical Example*

In this section, a simple numerical example is given to demonstrate the quantization error performance of the presented sub-optimal method compared to the conventional linear average based WB amplitude.

For the example in Section II.B, where two observed SB amplitudes of a beam are $(0.5, 1)$, based on the method presented in previous section, the sub-optimal WB amplitude $p_{l,i}^{(\text{Sub-WB})}$ shall be 0.9 instead of 0.75 which is the linear average amplitude over 2 SBs. Accordingly, the sub-optimal SB amplitude quantizer are [0.45, 0.9]. As a result, the quantized SB amplitude is [0.45 0.9], and the resulted RNSQE can be calculated as

$$\varepsilon_Q = \sqrt{\frac{|0.5 - 0.45|^2 + |1 - 0.9|^2}{0.5^2 + 1}} = 0.1$$

It is clear that the sub-optimal method significantly outperforms the linear average based WB amplitude method. Specifically, RNSQE in the example has been reduced by 60%.

## V. SIMULATION RESULTS

In this section, more simulation results are given to compare the provided methods, i.e., optimal and sub-optimal methods with the conventionl linear average based WB amplitude method.

It is clear that if the beam amplitude is quasi-flat over all SBs, i.e., frequency-flat channel, linear average based WB amplitude and optimal method shall be identical and achieve the minimum quantization error. As such, we focus on frequeny-selective channel where SB amplitudes considerably fluctuate. To this end, in the simulations, each SB amplitude is generated by a Gaussian random varible with certain variance. And greater variance imitates larger frequency selectivity.

In addition, to ensure always positive SB amplitude, a predefined minimum SB amplitude is given in each simulation, and all generated Gaussian SB amplitude is equally shifted so that the minimum SB amplitude is equal to the predefined value.

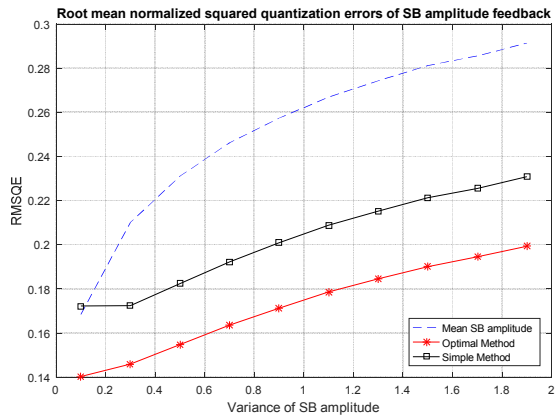

Fig. 3. Root Mean Normalized Squared Quantization Error Comparison, Minimum SB amplitude: 1.

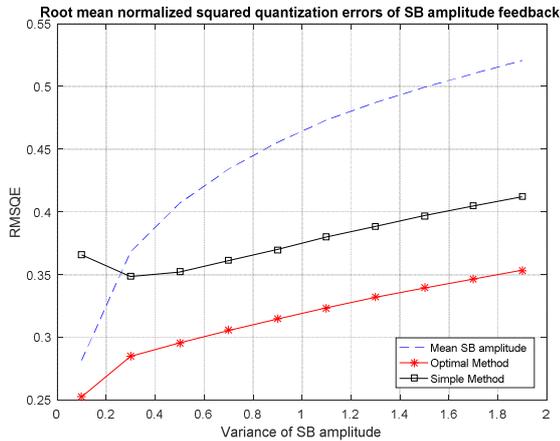

Fig. 4. Root Mean Normalized Squared Quantization Error Comparison, Minimum SB amplitude: 2

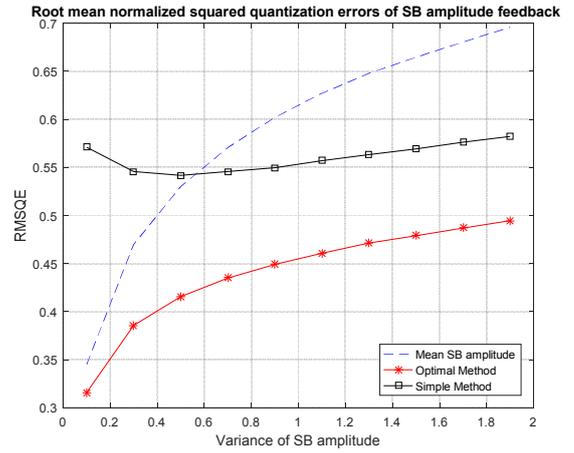

Fig. 5. Root Mean Normalized Squared Quantization Error Comparison, Minimum SB amplitude: 4

It is shown from Figs. 3-5 that presented optimal method achieves the best quantization performance among three methods in all frequency-selectivity regions. It is also observed that sub-optimal method outperforms the conventional method in high frequency-selective channels, i.e. large variance region. However, in frequency-flat channel, i.e., small variance region, sub-optimal method can be worse than the conventional method.

## VI. CONCLUSIONS

This paper presents Type-2 codebook based joint wideband and subband PMI feedback. In particular, two methods have been developed for WB amplitude calculation for Joint-WB-and-SB PMI feedback. The optimal method has been derived to achieve the minimum overall SB quantization errors. And sub-optimal method is also presented to reduce the computation complexity while achieving reasonably good quantization performance. Simulation results have demonstrated that optimal method can achieve the minimum quantization error all channels with different frequency-selectiveness, and in high frequency selective channel, sub-optimal method also outperforms the conventional linear average based WB amplitude method.

ACKNOWLEDGMENT

The research leading to these results received funding from the European Commission H2020 programme under grant agreement n°760809 (ONE5G project).

REFERENCES.

[1] 3GPP Technical Specification (TS) 38.211, "NR; Physical Channels and Modulation (Release 15)", December 2017.
[2] 3GPP Technical Specification (TS) 38.214, "NR; Physical Layer Procedure for Data (Release 15)", December 2017.
[3] Samsung etc., "WF on Type I and II CSI codebooks," R1-1709232, Hangzhou, China, May 15-19, 2017.